\documentclass[aps,prb,reprint,superscriptaddress,amsmath,amssymb,floatfix]{revtex4-2}

\usepackage{graphicx}
\usepackage{bm}
\usepackage{booktabs}
\usepackage{xcolor}
\usepackage{tikz}
\usetikzlibrary{arrows.meta,decorations.pathmorphing}
\usepackage[colorlinks=true,linkcolor=blue!60!black,citecolor=red!60!black,urlcolor=blue!70!black]{hyperref}

\graphicspath{{figures/}}

\newcommand{\Ztwo}{\mathbb{Z}_2}
\newcommand{\calO}{\mathcal{O}}

\newcommand{\Ne}{N_e}
\newcommand{\Np}{N_+}
\newcommand{\Nm}{N_-}

\newcommand{\one}{\mathbf{1}_d}
\newcommand{\hD}{\hat\Delta}

\begin{document}
\setcounter{dbltopnumber}{2}
\renewcommand{\dbltopfraction}{0.95}
\renewcommand{\dblfloatpagefraction}{0.75}
\renewcommand{\topfraction}{0.95}
\renewcommand{\bottomfraction}{0.9}
\renewcommand{\textfraction}{0.05}
\renewcommand{\floatpagefraction}{0.75}

\title{The \texorpdfstring{$\Ztwo$}{Z2} twist line defect of the 3D Ising CFT on the fuzzy sphere\texorpdfstring{\\}{ }
from multi-target density matrix renormalization group}

\author{Jing-Hao Jin}
\affiliation{State Key Laboratory of Surface Physics and Department of Physics, Fudan University, Shanghai 200433, China}
\author{Yang Qi}
\affiliation{State Key Laboratory of Surface Physics and Department of Physics, Fudan University, Shanghai 200433, China}
\affiliation{Collaborative Innovation Center of Advanced Microstructures, Nanjing 210093, China}

\date{\today}

\begin{abstract}
The fuzzy-sphere regularization provides a powerful framework for studying three-dimensional (3D) conformal field theories (CFTs) through radial quantization, but conformal defects have so far been studied with it only by exact diagonalization (ED) of small systems. In this work we realize the $\Ztwo$ twist line defect of the 3D Ising CFT on the fuzzy sphere by imposing antiperiodic boundary conditions on the $\Ztwo$-odd fermion flavor, and solve it with a multi-target density matrix renormalization group (DMRG) algorithm that optimizes several low-lying states of each symmetry sector in a common matrix-product basis. This gives the defect spectrum at system sizes well beyond the reach of ED (up to $2s+1=25$, or $49$ spinless orbitals). The spectrum obeys the spin--parity selection rule of the twist line and organizes into conformal multiplets along the line. Using the displacement operator as an intrinsic ruler, its protected descendant as a test, and the bulk stress tensor as an independent cross-check, we identify the leading finite-size corrections and obtain controlled continuum extrapolations, which sharpen the dimensions of several known defect primaries and resolve four higher-lying ones. This study establishes the fuzzy sphere combined with multi-target DMRG as a framework for precision studies of conformal defects.
\end{abstract}

\maketitle

\section{Introduction}
\label{sec:intro}
Conformal field theories (CFTs) describe the universal long-distance physics of continuous phase transitions. In three dimensions (3D) the local data of the Ising CFT are now known to high precision from the numerical conformal bootstrap~\cite{Kos2014,SimmonsDuffin2017,Poland2019}. A CFT is characterized, however, not only by its local operators but also by its extended ones: boundaries, interfaces, and line defects carry universal data of their own~\cite{Billo2016}. A conformal line defect preserves $\mathrm{SL}(2,\mathbb R)\times\mathrm{SO}(2)$, hosts defect primaries labeled by a scaling dimension $\Delta$ and a transverse spin $j$, and couples to the bulk through the bulk-to-defect OPE. Such lines describe magnetic impurities and pinning fields in critical magnets~\cite{Cuomo2022,HuHeZhu2024defect}, static charges in gauge theories, and, as we discuss next, the boundaries of symmetry defects~\cite{GKSW2015,McGreevy2023,SchaferNameki2024,Shao2023}.

An internal global symmetry is a product of on-site unitaries, $U=\prod_x u_x$. Because $u_x$ commutes with every term of the Hamiltonian that does not involve $x$, the symmetry may be performed on a subregion $D$ alone: $U_D=\prod_{x\in D}u_x$ leaves the Hamiltonian untouched except for the terms that straddle the boundary $\partial D$, where it inserts a seam across which the field changes sign. In spacetime this seam sweeps out the \emph{symmetry defect}. It is topological, it may be left open, and its boundary is a defect of its own, the \emph{twist line}, which does not depend on the surface chosen to end on it. In 3D the $\Ztwo$ symmetry defect of the Ising CFT is a surface, and a loop linking the boundary crosses the surface an odd number of times, so the spin field $\sigma$ changes sign around the twist line. On the lattice the line bounds a sheet of frustrated bonds; in the Wegner dual $\Ztwo$ gauge theory it is a Wilson line~\cite{Wegner1971}. In 2D the symmetry defect is the $\Ztwo$ line $\eta$, which together with the Kramers--Wannier duality line forms the Ising fusion category of topological defect lines~\cite{PetkovaZuber2001,Frohlich2004,Chang2019}. The analogue of the twist line is its endpoint, an operator of the twisted sector~\cite{Aasen2016}: the disorder field $\mu$ ($\Ztwo$-even, spin $0$) or a Majorana fermion ($\Ztwo$-odd, spin $\pm\tfrac12$). The 3D counterpart is the spin--parity selection rule, $\Ztwo$-odd defect operators carrying half-integer transverse spin~\cite{Billo2013,Gaiotto2013}. The line admits no relevant or marginal deformation, so its infrared spectrum is fixed by the bulk CFT.

Bill\`o et al.\ measured this spectrum in Monte Carlo simulations of the lattice model~\cite{Billo2013}, and Gaiotto et al.\ analyzed it with the $\epsilon$ expansion and the numerical bootstrap~\cite{Gaiotto2013}. Together they determined the leading odd primary $\psi$ and the even scalar $\mathcal O_s$ precisely [$\Delta_\psi=0.9187(6)$, $\Delta_{\mathcal O_s}=2.27(1)$], the spin-$\tfrac32$ primary to $3\%$, and the rest of the low-lying spectrum to $6$--$16\%$ (statistical errors only). Monodromy defects of the $\mathrm O(N)$ models have since been studied perturbatively and with the bootstrap~\cite{Soderberg2018,Bianchi2021,Giombi2022,GimenezGrau2022}, but nonperturbative data for the Ising twist line have advanced little. The line also matters in quantum magnets: through Wegner duality the gauged Ising (Ising$^*$) CFT describes the confinement transition out of a $\Ztwo$ spin liquid~\cite{FradkinShenker1979,Kitaev2003,SavaryBalents2017}, where a static anyon that braids nontrivially with the condensing one acts as a twist line. The resulting twisted sectors appear in finite-size torus spectra~\cite{Schuler2016,WhitsittSachdev2016}, and the disorder operator and its higher-form symmetry have been measured by quantum Monte Carlo~\cite{Zhao2021}; entanglement diagnostics based on flux threading probe related physics in spin liquids~\cite{Zhu2018}. Networks of twist lines encode the categorical data of the symmetry in the symmetry-topological-field-theory picture~\cite{Apruzzi2023,Kaidi2023,FreedMooreTeleman2024,ChatterjeeWen2023,Ji2024}, and tensor-network methods reach the same data by coarse-graining the partition function~\cite{LevinNave2007,GuWen2009,EvenblyVidal2015,Xie2012,LyuKawashima2025,GuoWei2024} or by constructing the fixed-point tensor directly~\cite{Cheng2025,Chen2024,Bao2025}, with topological defects included~\cite{Hauru2016}.

The fuzzy-sphere regularization realizes 3D CFTs with interacting fermions in the lowest Landau level (LLL) of a sphere around a monopole of strength $2s$; by the state--operator correspondence the energy spectrum gives the scaling dimensions~\cite{Zhu2023,HeZhu2026review}, and it has since produced OPE coefficients~\cite{HuHeZhu2023OPE}, four-point correlators~\cite{Han2023correlator}, the entropic $F$-function~\cite{Hu2025Ffunction}, and data for many other CFTs~\cite{Han2024,ZhouWangHe2026}. Magnetic line defects are realized by a local pinning field~\cite{HuHeZhu2024defect,Zhou2024gfunction} and boundaries by a fuzzy hemisphere or a spatial cut~\cite{Dedushenko2024,ZhouZou2025surface}, but the twist line is the edge of a symmetry defect and cannot be produced by any local term: it must be built into the boundary conditions, here by threading an Aharonov--Bohm $\pi$ flux seen only by the $\Ztwo$-odd fermions. Exact diagonalization (ED) is then limited by the Hilbert space, $\binom{4s+1}{2s+1}\sim2^{4s}$ at filling $N_e=2s+1$: about $10^9$ at $2s+1=17$ and $6\times10^{13}$ at $2s+1=25$. The defect makes matters worse, since it reduces $\mathrm{SO}(3)$ to $\mathrm{O}(2)$, so total angular momentum no longer organizes the spectrum into $\mathrm{SO}(3)$ multiplets~\cite{HeZhu2026review,Lauchli2025}.

The density matrix renormalization group (DMRG)~\cite{White1992,Schollwock2011} mitigates the Hilbert-space bottleneck. The LLL orbitals ordered by $m$ form a chain whose cut is a circle of latitude, so by the area law the entanglement of low-lying states grows only as $R\propto\sqrt s$, and matrix-product-state calculations have extended fuzzy-sphere studies of the Ising, $\mathrm O(2)$, and $\mathrm O(3)$ CFTs to sizes far beyond ED~\cite{Lauchli2025,Dey2026O2,Dey2026O3}. Spectroscopy, however, needs several excited states per sector, and sequential deflation accumulates errors and loses orthogonality. Multi-target DMRG~\cite{Baker2024,Hao2026} optimizes a bundle of low-lying states in a common matrix-product basis and solves this problem for the bulk CFT. Here we apply it to the twisted sphere. The broken rotation symmetry makes targeting several states per sector essential, because each transverse-spin sector holds several primaries interleaved with line descendants. We reach $2s+1=25$, use the displacement operator $D$ as an intrinsic ruler and its protected descendant $\partial_\parallel D$ as a test, show that, as expected from conformal perturbation theory~\cite{LaoRychkov2023}, the finite-size corrections are dominated by the leading irrelevant defect scalar $\mathcal O_s$, and extrapolate the spectrum to the continuum limit. This sharpens the dimensions of the less precisely known primaries ($\psi_{3/2}$, $t_+$, $D^o$, and $p^o$), resolves four higher-lying ones, and is cross-checked by the bulk stress tensor on the untwisted sphere.

The paper is organized as follows. Section~\ref{sec:model} describes the model, its orbitals and symmetry sectors; Sec.~\ref{sec:method} the algorithm, the calibration and the separation of primaries from descendants; Sec.~\ref{sec:results} the spectrum and its scaling analysis; and Sec.~\ref{sec:summary} the summary and outlook. Appendices~\ref{app:matrix}--\ref{app:budget} collect the technical details.

\section{Twist line on the fuzzy sphere}
\label{sec:model}

\subsection{Fuzzy-sphere model and twisted boundary condition}
\label{sec:tbc}
We formulate the critical 3D Ising CFT via the fuzzy-sphere Hamiltonian of Ref.~\cite{Zhu2023}. Spin-$\tfrac12$ fermions $\mathbf c_m=(c_{m\uparrow},c_{m\downarrow})^T$ reside on a 2-sphere of radius $R=\sqrt{s}$ (in units of the magnetic length $\ell_B=1$) threaded by $2s$ magnetic monopole flux quanta ($4\pi s$). The single-particle states are the LLL monopole harmonics~\cite{WuYang1976}
\begin{equation}
\Phi_{s,m}(\theta,\phi)=\mathcal N_{s,m}\,e^{i(s+m)\phi}
\sin^{s-m}\!\tfrac{\theta}{2}\,\cos^{s+m}\!\tfrac{\theta}{2},
\label{eq:LLL}
\end{equation}
with $m=-s,\dots,s$ the eigenvalue of the gauge-covariant canonical angular momentum $L_z=-i\partial_\phi-s$, and normalization $\mathcal N_{s,m}^2=\Gamma(2s+2)/[4\pi\,\Gamma(s-m+1)\Gamma(s+m+1)]$. The microscopic Hamiltonian combines short-range density interactions with a transverse magnetic field:
\begin{multline}
H=\int\! d\Omega_{1}d\Omega_2\,U(\Omega_{12})\big[n^0(\Omega_1)n^0(\Omega_2)
-n^z(\Omega_1)n^z(\Omega_2)\big]\\
-h\!\int\! d\Omega\, n^x(\Omega),
\label{eq:H}
\end{multline}
where $n^\alpha(\Omega)=\bm\Psi^\dagger\sigma^\alpha\bm\Psi$ ($\sigma^0=\mathbb 1_2$) denotes the projected density of the field $\bm\Psi(\Omega)=\sum_m\Phi_{s,m}(\Omega)\mathbf c_m$. The interaction kernel $U(\Omega_{12})$ is specified by Haldane pseudopotentials $(V_0,V_1)=(4.75,1)$~\cite{Haldane1983}. The $n^zn^z$ interaction drives Ising ferromagnetism, which is melted by the transverse field $h$, with the bulk Ising (Wilson--Fisher) critical point located at $h_c\simeq3.16$~\cite{Zhu2023}. The global $\Ztwo$ spin-flip symmetry acts as $\mathbf c_m\to\sigma^x\mathbf c_m$, mapping $n^z\to-n^z$.

Under radial quantization, a straight line defect piercing the coordinate origin intersects the spatial $S^2$ at the antipodal north (N) and south (S) poles [Fig.~\ref{fig:setup}(a)]. The twist line is defined by its monodromy: any $\Ztwo$-odd bulk operator changes sign upon winding around the line. We realize this condition by applying a local Hadamard rotation to diagonalize the $\Ztwo$ symmetry, the same change of basis that is used to exploit a flavor symmetry in bulk DMRG~\cite{Hao2026}: $c_{m,\pm}=(c_{m\uparrow}\pm c_{m\downarrow})/\sqrt2$. In this $(+,-)$ flavor basis, $\sigma^x\to\tau^z$ and $\sigma^z\to\tau^x$ (with $\tau^\alpha$ the flavor Pauli matrices), such that the Ising parity simply measures flavor parity:
\begin{equation}
\mathcal C=(-1)^{\Nm},\qquad \Nm=\textstyle\sum c^\dagger_{-}c_{-}.
\label{eq:C}
\end{equation}
The $(+)$ and $(-)$ fermion flavors carry $\Ztwo$ parity $+1$ and $-1$, respectively. The twist line is then engineered without modifying the local operator density, but purely by imposing twisted boundary conditions across a branch-cut meridian connecting the two punctures:
\begin{equation}
\Psi_\pm(\theta,\phi+2\pi)=\pm\,\Psi_\pm(\theta,\phi).
\label{eq:TBC}
\end{equation}
Equivalently, the odd flavor experiences an Aharonov--Bohm flux of $\pi$ threaded through the twist line, while the even flavor is unaffected. Because $H$ contains $\Psi_-$ exclusively in bilinear pairs, the Hamiltonian density remains single-valued, ensuring that the defect manifests entirely through the altered Hilbert space on which $H$ operates.

\subsection{Orbital geometry and filling}
\label{sec:orbitals}
\begin{figure}[htb]
\centering
\resizebox{\columnwidth}{!}{%
\begin{tikzpicture}[>=Latex,scale=0.95]
\begin{scope}
\shade[ball color=blue!8,opacity=0.9] (0,0) circle (1.25);
\draw[thick] (0,0) circle (1.25);
\draw[gray,densely dashed] (-1.25,0) arc (180:0:1.25 and 0.32);
\draw[gray] (-1.25,0) arc (180:360:1.25 and 0.32);
\draw[very thick,red!70!black] (0,-1.85)--(0,1.85);
\fill[red!70!black] (0,1.25) circle (2pt) node[right=2pt,font=\scriptsize] {N};
\fill[red!70!black] (0,-1.25) circle (2pt) node[right=2pt,font=\scriptsize] {S};
\draw[ultra thick,orange!85!black,decorate,decoration={snake,amplitude=0.6pt,segment length=4pt}]
  (0,1.25) arc (90:270:0.55 and 1.25);
\node[font=\scriptsize,orange!85!black,align=center] at (-1.35,-1.55) {branch cut\\$\Psi_-\to-\Psi_-$};
\node[font=\scriptsize,red!70!black] at (0.55,1.75) {defect};
\node[font=\small] at (-1.6,1.6) {(a)};
\end{scope}
\begin{scope}[xshift=3.1cm]
\node[font=\small] at (-0.35,1.6) {(b)};
\draw[thick,blue!70!black] (0,0.8)--(4.4,0.8);
\foreach \x in {0.2,0.9,1.6,2.3,3.0,3.7,4.4}{\fill[blue!70!black] (\x-0.0,0.8) circle (2.3pt);}
\draw[thick,red!70!black] (0,-0.2)--(4.4,-0.2);
\foreach \x in {0.55,1.25,1.95,2.65,3.35,4.05}{\fill[red!70!black] (\x,-0.2) circle (2.3pt);}
\node[font=\scriptsize,blue!70!black,anchor=south] at (2.3,0.95) {$(+)$: $m=-s,\dots,s$ \;($2s+1$)};
\node[font=\scriptsize,red!70!black,anchor=north] at (2.3,-0.35) {$(-)$: $\tilde m=-s+\tfrac12,\dots,s-\tfrac12$ \;($2s$)};
\draw[<->,thick,gray] (0.2,0.3)--(0.55,0.3) node[midway,above=-1pt,font=\tiny] {$\tfrac12$};
\node[font=\scriptsize,align=center] at (2.3,-1.35) {$4s+1$ orbitals, $N_e=2s+1$\\[1pt]
$(\Ne,\,2L_z,\,(-1)^{N_-})$ conserved};
\end{scope}
\end{tikzpicture}}
\caption{Twist line on the fuzzy sphere. (a)~In radial quantization the
straight twist line (red) crosses the sphere at the north (N) and south (S)
poles. The $\Ztwo$ symmetry defect ending on the line meets the sphere along
a meridian (wavy line). Across it the odd flavor $\Psi_-$ changes sign and
the even flavor $\Psi_+$ is continuous. No term is added to the Hamiltonian.
(b)~Single-particle orbitals at $2s+1=7$. The even flavor keeps the integer
grid $m=-s,\dots,s$ ($2s+1$ orbitals). The antiperiodic odd flavor lives on
the half-integer grid shifted by $\tfrac12$ ($2s$ orbitals), and all its
orbitals vanish at the poles as $\theta^{1/2}$ or faster. The half-integer
grid is what forces $\Ztwo$-odd states to have half-integer $L_z$,
Eq.~\eqref{eq:spin-parity}. In the MPS the $4s+1$ orbitals are ordered by
$m$, so the two grids alternate.}
\label{fig:setup}
\end{figure}
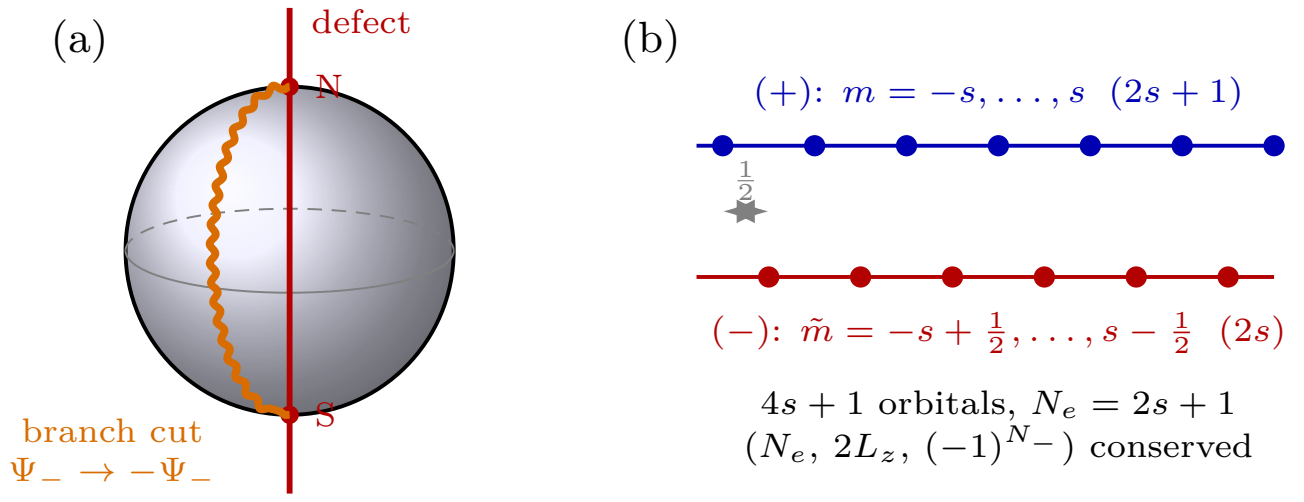

The boundary condition~\eqref{eq:TBC} reorganizes the single-particle orbital grid in two ways [Fig.~\ref{fig:setup}(b)]. First, the azimuthal phase factor $e^{i(s+m)\phi}$ in Eq.~\eqref{eq:LLL} restricts $m\in\mathbb Z$ for periodic fields, but shifts the odd flavor to a half-integer angular momentum grid, $\tilde m\in\mathbb Z+\tfrac12$. Second, analytic continuation of the monopole harmonics~\eqref{eq:LLL} to half-integers reveals the asymptotic polar profiles: $|\Phi_{s,\tilde m}|\propto\theta^{s-\tilde m}$ near the north pole and $|\Phi_{s,\tilde m}|\propto(\pi-\theta)^{s+\tilde m}$ near the south pole. Finiteness of the wave functions at both punctures enforces $|\tilde m|\le s$, which for half-integers yields $|\tilde m|\le s-\tfrac12$. The resulting single-particle spectrum comprises
\begin{equation}
\begin{aligned}
(+):&\quad m=-s,\dots,s, && N_{\rm orb,+}=2s+1,\\
(-):&\quad \tilde m=-s+\tfrac12,\dots,s-\tfrac12, && N_{\rm orb,-}=2s,
\end{aligned}
\label{eq:orbitals}
\end{equation}
totaling $N_{\rm orb}=4s+1$ orbitals. Every $(-)$ orbital vanishes at both punctures at least as $\theta^{1/2}$ and carries a half-integer azimuthal winding $e^{i\phi/2}$ relative to the $(+)$ flavor, matching the monodromy of a $\Ztwo$-odd field around the twist line. The $(+)$ orbitals with $m=\pm s$ remain finite at the poles, microscopically distinguishing the two flavors.

The twist carries no charge, so we keep the bulk filling $N_e=2s+1$. At large $h$ the ground state at this filling is the filled $(+)$ shell, with $L_z=0$ and $N_-=0$; at $h_c$ the ground state of the same sector is the defect identity $\one$. The alternative $N_e=2s$ leaves a hole in the $(+)$ shell. In ED this hole localizes at a puncture, where the $(-)$ orbitals vanish, and produces a degenerate ground state at $2L_z=\pm2s$ and low-lying spin-$1$ states incompatible with defect conformal symmetry. We therefore fix $N_e=2s+1$.

\subsection{Hamiltonian and symmetry sectors}
\label{sec:symmetry}
Projecting the normal-ordered interaction~\eqref{eq:H} onto the twisted basis~\eqref{eq:orbitals} yields the second-quantized Hamiltonian with twisted boundary conditions (TBC),
\begin{multline}
H^{\rm TBC}=-h(\Np-\Nm)+\frac14\sum_{\{m_i,\lambda_i\}}
(\lambda_2\lambda_3+\lambda_1\lambda_4)\\
\times V_{m_1m_2m_3m_4}\,\delta_{m_1+m_2,m_3+m_4}\,
c^\dagger_{m_1\lambda_1}c^\dagger_{m_2\lambda_2}c_{m_3\lambda_3}c_{m_4\lambda_4},
\label{eq:HTBC}
\end{multline}
where $\lambda_i=\pm1$ indexes the flavor, and each $m_i$ runs over the respective integer or half-integer grid. The flavor weight $(\lambda_2\lambda_3+\lambda_1\lambda_4)/2$ equals $+1$ for the density channel ($\lambda_1=\lambda_4$, $\lambda_2=\lambda_3$), $-1$ for flavor exchange ($\lambda_1=-\lambda_4$, $\lambda_2=-\lambda_3$), and vanishes otherwise, matching $n^0n^0-n^zn^z$ in the rotated basis. The matrix elements $V$ are determined by continuation of the Clebsch--Gordan coefficients to half-integers via Euler $\Gamma$ functions, exactly matching the real-space overlap integrals on the sphere (Appendix~\ref{app:matrix}).

Because the flavor-exchange process $c^\dagger_+c^\dagger_+c_-c_-$ scatters pairs between flavors, $\Np$ and $\Nm$ are not independently conserved. The Hamiltonian preserves the total particle number $\Ne$, the azimuthal angular momentum $L_z=\sum_m m\,n_{m+}+\sum_{\tilde m}\tilde m\,n_{\tilde m-}$, and the Ising parity $(-1)^{\Nm}$. Because every $(-)$ orbital carries half-integer $\tilde m$, the system satisfies an exact microscopic selection rule:
\begin{equation}
e^{2\pi iL_z}=(-1)^{\Nm}.
\label{eq:spin-parity}
\end{equation}
This is the spin--parity selection rule of the twist line~\cite{Billo2013,Gaiotto2013}: a $\Ztwo$-odd defect operator changes sign under a $2\pi$ rotation around the line and has half-integer transverse spin $j=L_z$. The $\pi$-rotation $R=e^{-i\pi L_y}$ exchanges the punctures and maps $L_z\to-L_z$, generating with $\mathrm{SO}(2)$ the spatial symmetry group $\mathrm{O}(2)$. It is the product $R=P_\perp S$ of a transverse reflection $P_\perp$ and the reversal of the line $S:x_\parallel\to-x_\parallel$ (the S-parity of Ref.~\cite{Billo2013}); hence levels with $\pm j$ are degenerate for $j\neq0$, and $j=0$ levels carry $R=P_\perp S=\pm1$. On the LLL a spatial reflection reverses the monopole flux and must be combined with particle--hole conjugation, which is broken at integer filling $N_e=2s+1$ because $N_{\rm orb}=4s+1$ is odd; $P_\perp$ and $S$ therefore emerge separately only at low energy, whereas the 3D rotation $R\in\mathrm{SO}(3)$ preserves the monopole flux and is an exact microscopic symmetry. Following Ref.~\cite{Billo2013}, S-odd operators carry the superscript $o$.

The twist line needs no tuning of defect couplings. A relevant deformation of a line defect must be an $\mathrm{SO}(2)$ scalar with $\Delta<1$. By Eq.~\eqref{eq:spin-parity} every $j=0$ defect operator is $\Ztwo$-even, and the lowest nontrivial one, $\mathcal O_s$ with $\Delta\simeq2.27$, is irrelevant. Tuning the bulk field to $h_c$ therefore suffices for the line to flow to the conformal twist line.

\section{Method}
\label{sec:method}

\subsection{Multi-target DMRG on the twisted sphere}
\label{sec:dmrg}
DMRG represents states as matrix-product states (MPS), whose cost is set by the bond dimension $\chi$ needed for the entanglement across a cut rather than by the Hilbert-space dimension. Ordering the LLL orbitals by $m$ makes each cut of the chain a cut of the sphere along a circle of latitude. The entanglement then obeys an area law, $S\propto R\propto\sqrt{N_{\rm orb}}$, so $\chi$ grows as $e^{a\sqrt{N_{\rm orb}}}$ while the sector dimension grows as $e^{bN_{\rm orb}}$~\cite{Hao2026}.

Single-state DMRG obtains excited states by sequential deflation with penalty projectors, which accumulates errors and loses orthogonality. For the bulk model, Ref.~\cite{Hao2026} also adds a penalty $\lambda\mathbf L^2$ that selects the states with total angular momentum $l=|L_z|$. The twist line breaks $\mathrm{SO}(3)$ to $\mathrm{O}(2)$ (generated by rotations about the line and $R=e^{-i\pi L_y}$), so $\mathbf L^2$ is not conserved and no such penalty is available. Each sector $2L_z$ then contains several primaries interleaved with the line descendants of lower ones (Table~\ref{tab:assign}), and resolving them requires several states per sector. In multi-target DMRG~\cite{Baker2024,Hao2026} the $K$ lowest states of a sector share all site tensors and differ only in the orthogonality center $\Theta_k$, $k=1,\dots,K$. In each two-site update the $K$ centers are optimized together by block Lanczos~\cite{Baker2024}, and the bond basis is truncated with the state-averaged density matrix $\rho=K^{-1}\sum_k\mathrm{Tr}_{\rm env}(\Theta_k\Theta_k^\dagger)$, followed at the end of each sweep by a $K\times K$ Gram--Ritz rotation that restores global orthonormality (Appendix~\ref{app:algo}).

For the twist line the two grids are interleaved by latitude, $m_a=-s+(a-1)/2$ for $a=1,\dots,4s+1$, so each of the $4s+1$ orbitals is a spinless fermionic site. As in the rotated bulk formulation~\cite{Hao2026}, flavor-pair scattering $c^\dagger_+c^\dagger_+c_-c_-$ breaks the separate conservation of $N_+$ and $N_-$, leaving $N_e$, $2L_z$, and $(-1)^{N_-}$ as the conserved Abelian charges. We build the MPS and the matrix-product operator (MPO) with ITensor~\cite{ITensor} and FuzzifiED~\cite{Zhou2025FuzzifiED}, and use White's density-matrix perturbation~\cite{White2005} in the first sweeps.

Starting from random MPS, all $K=4$ targets reproduce our own ED of the twisted Hamiltonian to $1.1\times10^{-13}$ at $2s+1=5$ and $7$ (Appendix~\ref{app:conv}), which benchmarks the targets before the calibrated $\hD_\psi=0.9192$ [Eq.~\eqref{eq:ruler} below] is used at larger sizes. We then compute, at $(V_0,V_1)=(4.75,1)$ and $h=3.16$, the sizes $2s+1=17,21,25$ ($33$ to $49$ orbitals), with $K=4$ states in each sector $2L_z=1,\dots,4$ and the vacuum at $2L_z=0$, at bond dimensions up to $\chi=3703$. Energy changes between the last sweeps are below $10^{-7}$, and discarded weights are about $10^{-7}$ (Appendix~\ref{app:conv}), far below the finite-size drift of the levels. The $j=0$ sector, which holds the vacuum together with $\mathcal O_s$, $p^o$, and $\partial_\parallel\mathcal O_s$, is computed separately with $K=4$ at $2s+1=17$, $21$, and $25$; the run at $2s+1=25$ has completed two sweeps and is treated as preliminary (Appendix~\ref{app:conv}). From $2s+1=17$ to $25$ the sector dimension grows by a factor of $3\times10^4$, to $5.0\times10^{11}$ ($4$~TB for one vector), whereas $\chi$ grows by a factor of $2$--$3$ and the time per sweep by $7$ (Fig.~\ref{fig:cost}). The assignment of primaries and line descendants is described in Sec.~\ref{sec:lambda}.

\subsection{Displacement calibration}
\label{sec:calibration}
\begin{figure*}[tp]
\centering
\includegraphics[width=\textwidth]{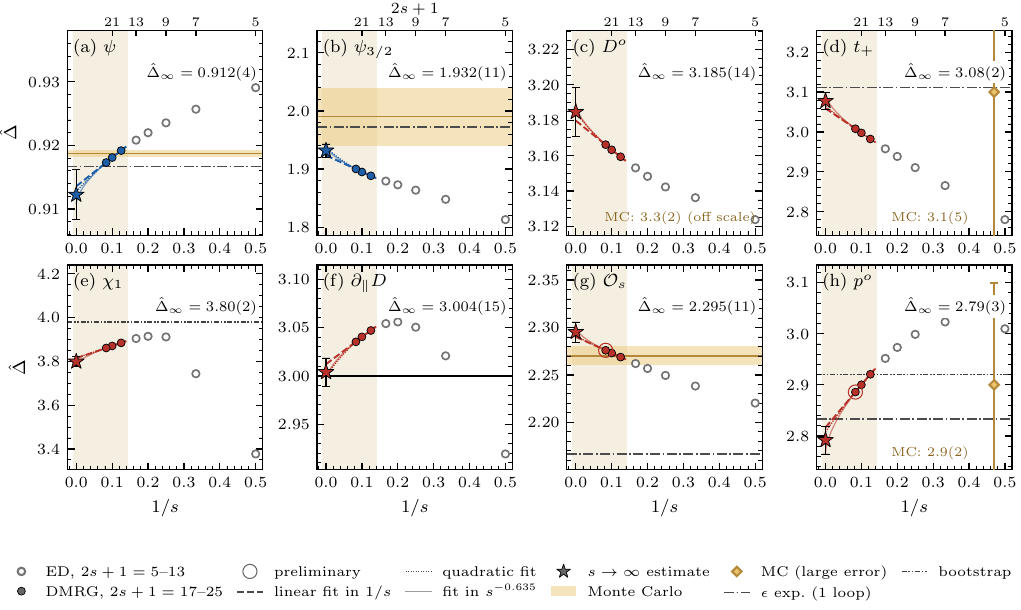}
\caption{Approach to the continuum limit. Calibrated dimensions versus $1/s$
(top axis: $2s+1$). Open gray circles: ED, $2s+1=5$--$13$. Filled circles on
the shaded background: multi-target DMRG, $2s+1=17$--$25$. Lines: fits to the
three DMRG points, linear (dashed) and quadratic (dotted) in $1/s$ and linear in
$s^{-0.635}$ (thin solid) [Eq.~\eqref{eq:omega}]. Star and number in each panel:
infinite-size estimate with the error of Appendix~\ref{app:budget}. Yellow band: MC
value~\cite{Billo2013} (quoted in the panel when off scale);
dash-dotted: one-loop $\epsilon$ expansion; dash-double-dotted: spin-$1$
bootstrap estimate~\cite{Gaiotto2013}. (a)~$\psi$, (b)~$\psi_{3/2}$,
(c)~$D^o$, (d)~$t_+$, (e)~$\chi_1$, (f)~the protected descendant
$\partial_\parallel D$, whose exact dimension $3$ is the solid line,
(g)~$\mathcal O_s$, (h)~$p^o$. In (g) and (h) the large ring marks the
preliminary point at $2s+1=25$.}
\label{fig:extrap}
\end{figure*}

Energies on the fuzzy sphere give scaling dimensions up to a non-universal velocity $v$, $E_\calO-E_{\one}=v\Delta_\calO/R$. We fix $v$ with the displacement operator $D$, the lowest $\Ztwo$-even primary at $|j|=1$, whose dimension $\Delta_D=2$ follows from the Ward identity of transverse translations:
\begin{equation}
\hD_\calO=2\,\frac{E_\calO-E_{\one}}{E_D-E_{\one}}.
\label{eq:ruler}
\end{equation}
Since $D$ belongs to the twisted system itself, no separate untwisted calculation enters, unlike for the bulk stress-tensor ruler ($\Delta_T=3$). The protected descendant $\partial_\parallel D$ ($\Delta=3$) then tests the calibration and the extrapolation together; it extrapolates to $3.004(15)$ [Fig.~\ref{fig:extrap}(f), Sec.~\ref{sec:continuum}].

\begin{figure*}[tp]
\centering
\includegraphics[width=\textwidth]{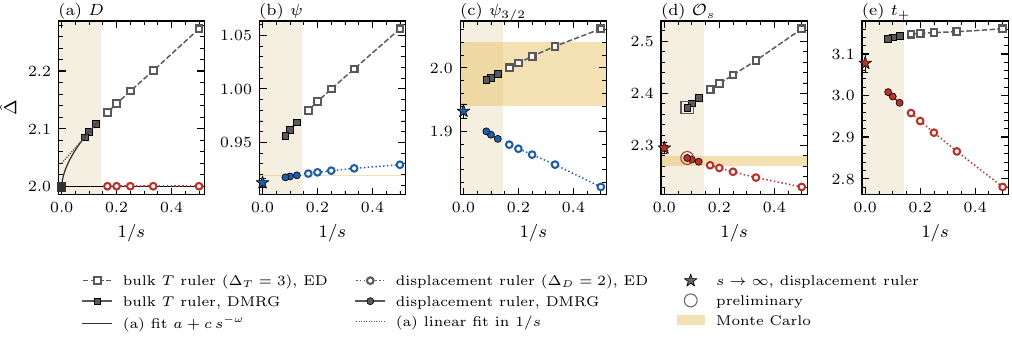}
\caption{Displacement ruler versus bulk stress-tensor ruler for (a)~$D$,
(b)~$\psi$, (c)~$\psi_{3/2}$, (d)~$\mathcal O_s$ and (e)~$t_+$, plotted versus $1/s$.
Squares: dimensions calibrated with the bulk stress tensor ($\Delta_T=3$),
open for ED ($2s+1=9$--$13$) and filled for DMRG ($2s+1=17$--$25$, shaded).
Circles: displacement ruler ($\Delta_D=2$), open for ED and filled for DMRG.
In (a) the solid line is the free fit $2(a+c\,s^{-\omega})$ through the three DMRG
squares and the filled square at $1/s=0$ its value $2a=1.999$; the dotted line
is a linear fit in $1/s$ to the same points. In (b)--(e) the stars are the
displacement-ruler estimates of Table~\ref{tab:results}. Large rings mark the
preliminary $j=0$ point at $2s+1=25$. The light band marks the MC value where
its error fits the plotted range. The gap between squares and circles measures
how far the twisted and untwisted systems are from a common conformal limit at
the same size; it closes in the continuum to within about $2\%$
(Sec.~\ref{sec:calibration}).}
\label{fig:rulers}
\end{figure*}

Figure~\ref{fig:rulers} tests the calibration against an independent ruler,
the bulk stress tensor ($\Delta_T=3$). Its gap $\Delta E_T$ is measured on an
untwisted sphere of the same size, with ED at $2s+1=9$--$13$ and with DMRG at
$2s+1=17$--$25$ (Appendix~\ref{app:conv}). If the twisted and the untwisted
systems were already conformal with a common velocity, the two rulers would
agree. They differ by the factor
$v_D/v_T=\tfrac32(E_D-E_{\one})/\Delta E_T$, which falls from $1.054$ at
$2s+1=17$ to $1.043$ at $2s+1=25$, so at finite size the bulk ruler gives larger
dimensions for every operator. The ED sizes alone suggested that this offset
survives the extrapolation: a linear fit in $1/s$ to $2s+1=9$--$13$ gives
$\Delta_D=2.05$ on the bulk ruler. The DMRG sizes show instead that the ratio
bends toward one. A free power law $v_D/v_T=a+c\,s^{-\omega}$ through the three
DMRG points gives $a=0.999$ and $\omega=0.58$, and reproduces the three ED
points to $2\times10^{-3}$. The
fit forms of Appendix~\ref{app:budget} give $a=1.004$--$1.020$, and fits with
the bulk exponent $0.415$ or with $s^{-1/2}$ give $0.98$ and $0.99$. The two
rulers therefore share their continuum limit to within about $2\%$. The fitted
exponent lies between the bulk ($0.415$) and defect ($0.635$) correction exponents
of Eq.~\eqref{eq:omega}. Bulk corrections enter the twisted and the untwisted
energies alike and may partly cancel in the ratio. The same free fit applied to the
bulk-ruler data for $\psi$, $D^o$, and $\partial_\parallel D$ gives $0.911$,
$3.19$, and $2.95$, within about $2\%$ of the displacement-ruler results.
We keep the displacement ruler for all results, because it is protected and is
measured in the same system as the defect levels, and use the bulk ruler as a
consistency check rather than as an error estimate. The residual error of the
displacement ruler is measured directly on $\partial_\parallel D$
(Appendix~\ref{app:budget}).


\subsection{Primaries and line descendants}
\label{sec:lambda}
Within a sector $(|j|,\Ztwo)$ a line descendant $\partial_\parallel\calO$
has the same quantum numbers as further primaries at similar $\hD$, so the
assignment needs more than the energies. We use the $\mathrm{SL}(2,\mathbb R)$
generator $P_\parallel+K_\parallel$, which on the sphere is represented by
$\Lambda_z=\int d\Omega\,\cos\theta\,\mathcal H(\Omega)$, the $\ell=1$ component of the
energy density along the defect axis~\cite{Fan2024,Fardelli2025}. In the
continuum it connects only neighboring members of one
$\mathrm{SL}(2,\mathbb R)$ multiplet,
\begin{equation}
\langle\partial_\parallel^{k+1}\calO|P_\parallel+K_\parallel|\partial_\parallel^k\calO\rangle=\sqrt{(k+1)(2\Delta_\calO+k)},
\label{eq:PK}
\end{equation}
for unit-normalized states $|\partial_\parallel^k\calO\rangle\propto P_\parallel^k|\calO\rangle$, and vanishes between different multiplets. We compute the matrix of
$\Lambda_z$ in the ED eigenstates of each sector, scale it by the same factor
that converts energies to $\hD$, and divide each element by the right-hand
side of Eq.~\eqref{eq:PK}, evaluated with the $\hD$ of the lower state. The resulting ratio
$r$ is of order one for a descendant pair and small otherwise. A level is
assigned as a descendant of the state with which it has the largest ratio if
$r>0.5$, and as a primary otherwise.

Table~\ref{tab:lambda} lists the ratios at $2s+1=9$, $11$, $13$. Every
descendant pair has $r=0.71$--$0.86$, and the value grows with size in each
tower. The common factor below one reflects the normalization of $\Lambda_z$
on a finite sphere, which we do not fix. The primaries have $r\le0.09$ at
$2s+1=13$, an order of magnitude smaller. The exceptions are pairs that are
nearly degenerate at small size: $\chi_1$ has $r=0.39$ at $2s+1=9$, and at
$2s+1=11$ the level-one state of $t_+$ and the primary $\chi_2$ lie $0.012$
apart and share the matrix element ($r=0.65$ and $0.40$). At $2s+1=13$ the two
levels have crossed and separate cleanly again ($0.79$ and $0.13$); this
crossing is why $\chi_2$ lies below $\partial_\parallel t_+$ at the DMRG sizes.
In the same $|j|=2$ sector, the fourth level $\chi_2'$ ($\hD\simeq4.14$ at $2s+1=25$) is also a primary, since the only lower states are $t_+$, $\chi_2$, and $\partial_\parallel t_+$, whereas the next descendants $\partial_\parallel\chi_2$ and $\partial_\parallel^2 t_+$ lie near $\Delta\gtrsim4.8$.

In the $j=0$ sector the rotation $R=e^{-i\pi L_y}$, which exchanges the
punctures ($\theta\to\pi-\theta$), is a symmetry,
and its eigenvalues are $\pm1$ to $10^{-15}$. Since $\Lambda_z$ is odd under
$R$, a level-one descendant has the opposite $R$ of its primary. The $j=0$ primaries
$\mathcal O_s$ ($P_\perp=S=+1$) and $p^o$ ($P_\perp=S=-1$) both carry $R=P_\perp S=+1$, their first descendants carry $R=-1$, and $\Lambda_z$
between $\mathcal O_s$ and $p^o$ vanishes by $R$-symmetry; this is the basis of the
identification of $p^o$ as a primary in Sec.~\ref{sec:primaries}.

These diagnostics use ED states. The DMRG sizes contain the same levels in
the same order within every sector, and each DMRG level is identified with
the ED level it connects to continuously (Fig.~\ref{fig:Ds}). This holds also
at $j=0$, where the three lowest excited DMRG levels at $2s+1=17$--$25$
continue $\mathcal O_s$, $p^o$, and $\partial_\parallel\mathcal O_s$ from ED
without crossings; our DMRG does not resolve $R$, so this ordering is the
identification. Evaluating $\Lambda_z$ and $R$ directly in the DMRG bundles
requires their MPOs and is left for future work.

\begin{table}[htb]
\caption{Normalized $\Lambda_z$ matrix elements $r$ (see text) from ED at
$2s+1=9,11,13$. Upper block: descendant pairs, $r$ between
$\partial_\parallel^{n}\calO$ and $\partial_\parallel^{n-1}\calO$; the
continuum value, up to the overall normalization of $\Lambda_z$, is the same
for all pairs. Lower block: primaries, largest $r$ with any lower level of
the sector. $\hD$ is the calibrated value at $2s+1=13$.}
\label{tab:lambda}
\begin{ruledtabular}
\begin{tabular}{llcccc}
pair or level & $(|j|,\Ztwo)$ & $\hD$ & $9$ & $11$ & $13$\\
\hline
$\psi\to\partial\psi$ & $(\tfrac12,-)$ & $1.995$ & $0.751$ & $0.779$ & $0.799$\\
$\partial\psi\to\partial^2\psi$ & $(\tfrac12,-)$ & $3.076$ & $0.793$ & $0.833$ & $0.859$\\
$\psi_{3/2}\to\partial\psi_{3/2}$ & $(\tfrac32,-)$ & $2.924$ & $0.763$ & $0.805$ & $0.834$\\
$D\to\partial D$ & $(1,+)$ & $3.054$ & $0.741$ & $0.775$ & $0.795$\\
$t_+\to\partial t_+$ & $(2,+)$ & $3.924$ & $0.706$ & $0.651$ & $0.789$\\
$\mathcal O_s\to\partial\mathcal O_s$ & $(0,+)$ & $3.268$ & $0.739$ & $0.778$ & $0.805$\\
$p^o\to\partial p^o$ & $(0,+)$ & $4.057$ & $0.771$ & $0.816$ & $0.844$\\
\hline
$\chi_{1/2}$ & $(\tfrac12,-)$ & $3.635$ & $0.044$ & $0.023$ & $0.024$\\
$\chi_{3/2}$ & $(\tfrac32,-)$ & $3.270$ & $0.042$ & $0.044$ & $0.042$\\
$D^o$ & $(1,+)$ & $3.153$ & $0.042$ & $0.045$ & $0.089$\\
$\chi_1$ & $(1,+)$ & $3.904$ & $0.391$ & $0.114$ & $0.071$\\
$\chi_2$ & $(2,+)$ & $3.875$ & $0.029$ & $0.398$ & $0.133$\\
$p^o$ & $(0,+)$ & $2.952$ & $0$ & $0$ & $0$\\
\end{tabular}
\end{ruledtabular}
\end{table}

\section{Results}
\label{sec:results}

\subsection{Defect spectrum and conformal multiplets}
\label{sec:spectrum}
\begin{figure}[htb]
\centering
\includegraphics[width=\columnwidth]{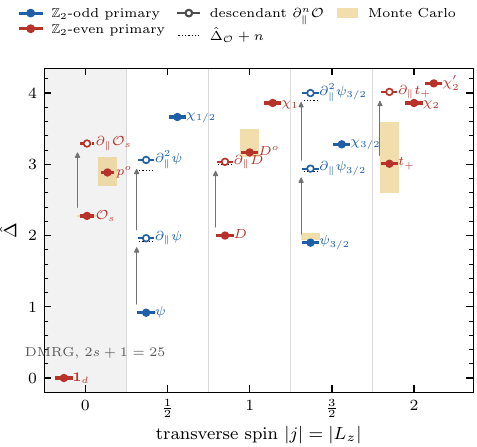}
\caption{Defect spectrum at $2s+1=25$ ($49$ orbitals) from multi-target
DMRG, calibrated with $\hD_D=2$ [Eq.~\eqref{eq:ruler}], versus the transverse
spin $|j|=|L_z|$. Blue (red): $\Ztwo$-odd (even) states. Filled circles:
primaries; open circles: line descendants $\partial_\parallel^n\calO$, linked
to their primary by arrows; dotted bars: conformal positions $\hD_\calO+n$.
Shaded rectangles: MC values~\cite{Billo2013}. Gray $|j|=0$
column (dashed bars): separate $K=4$ run of this sector (preliminary,
Appendix~\ref{app:conv}).}
\label{fig:spectrum}
\end{figure}
Figure~\ref{fig:spectrum} shows the calibrated spectrum at $2s+1=25$ ($49$ orbitals). Every $\Ztwo$-odd level (blue) sits at half-integer $|j|$ and every even level (red) at integer $|j|$, as required by Eq.~\eqref{eq:spin-parity}. The lowest odd level is the spin-$\tfrac12$ primary $\psi$ at $\hD=0.917$, the leading term in the defect OPE of the bulk spin field, $\sigma(x)\supset b_{\sigma\psi}|x_\perp|^{\Delta_\psi-\Delta_\sigma}e^{i\phi/2}\psi(x_\parallel)+\text{H.c.}$; its spin-$\tfrac32$ partner $\psi_{3/2}$ is at $1.900$. Both lie about $0.1$ below the naive estimates $\Delta_\sigma+|j|$.

Along the line the conformal group is $\mathrm{SL}(2,\mathbb R)$, and each primary $\calO$ heads a tower of descendants $\partial_\parallel^n\calO$ with the same $|j|$ and $\Ztwo$ parity, at $\hD_\calO+n$ in the continuum. In Fig.~\ref{fig:spectrum} the towers are marked by arrows and the conformal positions by dotted bars. Table~\ref{tab:assign} lists the resulting assignment of the four lowest levels in every sector. Sectors forbidden by Eq.~\eqref{eq:spin-parity} are empty, each $|j|>0$ sector contains a primary with its first (and, for $\psi$ and $\psi_{3/2}$, second) line descendant, and the remaining levels are further primaries.

\begin{table}[htb]
\caption{Assignment of the four lowest levels in each symmetry sector $(2L_z,\Ztwo)$, in order of increasing energy at $2s+1=25$ (multi-target DMRG; for $L_z=0$ from the separate $K=4$ run). Sectors with $e^{2\pi iL_z}\neq(-1)^{\Nm}$, such as $(2L_z,\Ztwo)=(0,-)$ and $(1,+)$, contain no states by Eq.~\eqref{eq:spin-parity}. Here $\partial\equiv\partial_\parallel$; numerical values are in Table~\ref{tab:results}.}
\label{tab:assign}
\begin{ruledtabular}
\begin{tabular}{lcccc}
$(2L_z,\Ztwo)$ & 1st & 2nd & 3rd & 4th\\
\hline
$(0,+)$ & $\one$ & $\mathcal O_s$ & $p^o$ & $\partial\mathcal O_s$\\
$(1,-)$ & $\psi$ & $\partial\psi$ & $\partial^2\psi$ & $\chi_{1/2}$\\
$(2,+)$ & $D$ & $\partial D$ & $D^o$ & $\chi_1$\\
$(3,-)$ & $\psi_{3/2}$ & $\partial\psi_{3/2}$ & $\chi_{3/2}$ & $\partial^2\psi_{3/2}$\\
$(4,+)$ & $t_+$ & $\chi_2$ & $\partial t_+$ & $\chi_2'$\\
\end{tabular}
\end{ruledtabular}
\end{table}

\begin{table*}[tp]
\caption{Calibrated dimensions $\hD$ of twist-line operators. The size columns give $2s+1$, the number of orbitals of the even flavor;
$2s+1=13$ is ED and $2s+1=17,21,25$ are multi-target DMRG. ``Extrap.'' is the
infinite-size estimate from $2s+1=17$--$25$: the midpoint of linear and
quadratic fits in $1/s$ and a linear fit in $s^{-0.635}$, with the half-spread
of the three and the calibration error $0.40\%$ combined in quadrature
(Appendix~\ref{app:budget}). For the $j=0$ levels $\mathcal O_s$ (the lowest $\Ztwo$-even scalar primary, not to be confused with the sphere size $s$), $p^o$, and $\partial_\parallel\mathcal O_s$, the DMRG values are from separate $K=4$ runs, measured from the ground state of the same run; values marked $^\dagger$ are preliminary (Appendix~\ref{app:conv}). References: MC values from Table~2 of
Ref.~\cite{Billo2013} (for $D^o$ and $t_+$ the quoted error is statistical; an additional systematic error $0.3$ is estimated there);
``dev.'' is $(\hD_{\rm extrap}-\Delta_{\rm MC})/\sigma$ with the errors combined in quadrature, including the MC systematic error for $D^o$ and $t_+$. One-loop $\epsilon$ expansion at $\epsilon=1$ and bootstrap values of
Ref.~\cite{Gaiotto2013}.}
\label{tab:results}
\begin{ruledtabular}
\begin{tabular}{lccccccccl}
operator & $(|j|,\Ztwo)$ & $13$ & $17$ & $21$ & $25$ & extrap. & MC & dev. & $\epsilon$ exp.\ / bootstrap\\
\hline
$\psi$ & $(\tfrac12,-)$ & $0.9208$ & $0.9192$ & $0.9181$ & $0.9173$ & $0.912(4)$ & $0.9187(6)$ & $-1.6$ & $0.917$\\
$\psi_{3/2}$ & $(\tfrac32,-)$ & $1.8795$ & $1.8885$ & $1.8951$ & $1.9003$ & $1.932(11)$ & $1.99(5)$ & $-1.1$ & $1.972$\\
$\chi_{3/2}$ & $(\tfrac32,-)$ & $3.2696$ & $3.2748$ & $3.2780$ & $3.2804$ & $3.296(14)$ & --- &  & ---\\
$\chi_{1/2}$ & $(\tfrac12,-)$ & $3.6352$ & $3.6493$ & $3.6582$ & $3.6643$ & $3.705(18)$ & --- &  & ---\\
$\mathcal O_s$ & $(0,+)$ & $2.2620$ & $2.2688$ & $2.2730$ & $2.2759^\dagger$ & $2.295(11)$ & $2.27(1)$ & $+1.7$ & $2.167$\\
$p^o$ & $(0,+)$ & $2.9518$ & $2.9206$ & $2.9001$ & $2.8859^\dagger$ & $2.79(3)$ & $2.9(2)$ & $-0.5$ & $2.833$ / $2.92$\\
$D$ & $(1,+)$ & $2$ & $2$ & $2$ & $2$ & $2$ & $2$ &  & $2$ (Ward)\\
$D^o$ & $(1,+)$ & $3.1531$ & $3.1593$ & $3.1633$ & $3.1662$ & $3.185(14)$ & $3.3(2)$ & $-0.3$ & ---\\
$\chi_1$ & $(1,+)$ & $3.9045$ & $3.8845$ & $3.8704$ & $3.8612$ & $3.80(2)$ & --- &  & $3.98$ (bootstrap)\\
$t_+$ & $(2,+)$ & $2.9578$ & $2.9824$ & $2.9976$ & $3.0081$ & $3.08(2)$ & $3.1(5)$ & $0.0$ & $3.111$\\
$\chi_2$ & $(2,+)$ & $3.8750$ & $3.8695$ & $3.8645$ & $3.8609$ & $3.838(17)$ & --- &  & ---\\
$\chi_2'$ & $(2,+)$ & $4.0932$ & $4.1189$ & $4.1304$ & $4.1362$ & $4.17(3)$ & --- &  & ---\\
\hline
$\partial_\parallel\psi$ & $(\tfrac12,-)$ & $1.9954$ & $1.9810$ & $1.9702$ & $1.9621$ & $1.911(16)$ & $\Delta_\psi+1$ &  & \\
$\partial_\parallel D$ & $(1,+)$ & $3.0541$ & $3.0469$ & $3.0404$ & $3.0352$ & $3.004(15)$ & $3$ (exact) &  & \\
$\partial_\parallel\psi_{3/2}$ & $(\tfrac32,-)$ & $2.9240$ & $2.9371$ & $2.9395$ & $2.9392$ & $2.93(2)$ & $\Delta_{\psi_{3/2}}+1$ &  & \\
$\partial_\parallel t_+$ & $(2,+)$ & $3.9236$ & $3.9756$ & $4.0015$ & $4.0164$ & $4.11(5)$ & $\Delta_{t_+}+1$ &  & \\
$\partial_\parallel\mathcal O_s$ & $(0,+)$ & $3.2682$ & $3.2829$ & $3.2891$ & $3.2922^\dagger$ & $3.31(2)$ & $\Delta_{\mathcal O_s}+1$ &  & \\
\end{tabular}
\end{ruledtabular}
\end{table*}

At $2s+1=25$ the level-one spacings $\hD_{\partial_\parallel\calO}-\hD_\calO$ are $1.045$ ($\psi$), $1.035$ ($D$), $1.039$ ($\psi_{3/2}$), and $1.008$ ($t_+$); for $\psi$ the spacing decreases from $1.062$ at $2s+1=17$ (Fig.~\ref{fig:spacings}). At $j=0$ the spacing of $\mathcal O_s$ is $1.014$--$1.016$ at $2s+1=17$--$25$. The spacing of $D$ is protected, $\Delta_{\partial_\parallel D}-\Delta_D=1$, so $\hD_{\partial_\parallel D}-3$ directly measures the calibration error. It decreases from $0.047$ to $0.035$ over the DMRG sizes and extrapolates to $0.004(15)$.

Among the even levels, the spin-$2$ primary $t_+$ lies at $\hD=3.008$, about $0.2$ above $\Delta_\psi+\Delta_{\psi_{3/2}}\simeq2.82$, and $D$ at $2$, above $2\Delta_\psi\simeq1.83$. We further identify the S-odd vector $D^o$ at $3.166$, a second vector $\chi_1$ at $3.861$, and four primaries not reported in previous MC or ED studies, $\chi_{1/2}$, $\chi_{3/2}$, $\chi_2$, and $\chi_2'$ (Table~\ref{tab:results}).

\subsection{Continuum limit and scaling corrections}
\label{sec:continuum}
Figure~\ref{fig:extrap} shows the approach to the continuum limit, and Table~\ref{tab:results} lists the extrapolated values. Between $2s+1=17$ and $25$ every DMRG level moves by less than $0.06$ along a smooth, nearly linear curve in $1/s$, whereas the ED levels at $2s+1\le13$ curve away [Fig.~\ref{fig:extrap}(e,f)]: the smaller sizes are not yet in the scaling regime.

At finite size the fuzzy-sphere Hamiltonian is the fixed point perturbed by irrelevant operators. A bulk operator of dimension $\Delta_{\rm b}>3$ or a defect operator of dimension $\Delta_{\rm d}>1$ shifts the calibrated levels as
\begin{equation}
\hD(s)=\Delta+b\,s^{-(\Delta_{\rm b}-3)/2}+c\,s^{-(\Delta_{\rm d}-1)/2}+\dots,
\label{eq:omega}
\end{equation}
with $R\propto\sqrt{s}$ in units of the magnetic length. The leading bulk term comes from $\epsilon'$ ($\Delta\approx3.83$, exponent $0.415$), the leading defect term from $\mathcal O_s$ ($\Delta\approx2.27$, exponent $0.635$); higher operators and second-order terms give $s^{-1}$ and $s^{-0.83}$ (Appendix~\ref{app:budget}).

The protected $\partial_\parallel D$ tests this form. With $\Delta=3$ fixed, $b\,s^{-0.415}+c\,s^{-0.635}$ fits the three DMRG points with an rms residual of $1\times10^{-4}$, fifteen times smaller than the one-parameter fit $c/s$; the one-parameter fit $c\,s^{-0.635}$ alone still gives $5\times10^{-4}$, so the defect term dominates. A single power law with free exponent gives $\omega=0.70(2)$, between $0.635$ and $1$, so the subleading terms still matter at these sizes. We therefore extrapolate every level with three forms that bracket this exponent: linear and quadratic in $1/s$, and linear in $s^{-0.635}$. For $\partial_\parallel D$ they give $3.012$, $3.000$, and $2.996$ [Fig.~\ref{fig:extrap}(f)], so the ruler and the extrapolation together reproduce $\Delta=3$ to $0.4\%$. A residual error of the ruler rescales all levels by a common factor, so we take this as a relative calibration error $\sigma_{\rm cal}=0.0040\,\Delta$ for all operators and add it in quadrature to half the spread of the three fits, $\sigma_{\rm fit}$. The level-one spacings of $\psi$ and $\psi_{3/2}$ extrapolate linearly in $1/s$ to $1.011$ and $1.021$, and that of $\mathcal O_s$ to $1.021$, consistent with $\mathrm{SL}(2,\mathbb R)$ in the continuum (Appendix~\ref{app:spacings}).

\subsection{Defect primaries and continuum comparison}
\label{sec:primaries}
For $\psi$ the calibrated dimension decreases monotonically from $0.9235$ at $2s+1=9$ to $0.9173$ at $2s+1=25$ and passes through the MC band $0.9187(6)$~\cite{Billo2013} near $2s+1\approx19$ [Fig.~\ref{fig:extrap}(a)]. The three fits give $0.9136$, $0.9127$, and $0.9109$, so $\Delta_\psi=0.912(4)$; other forms ($s^{-1/2}$, $s^{-0.415}$, free exponent) give $0.907$--$0.911$ (Table~\ref{tab:forms}). This lies $1.6\sigma$ ($0.7\%$) below MC. On our side, possible sources are corrections beyond Eq.~\eqref{eq:omega} and the small finite-size drift of the bulk critical point ($h_c\simeq3.153$--$3.16$~\cite{Zhu2023,Hao2026}); on the MC side, systematic effects of the lattice geometry. A high-precision numerical bootstrap study of the twist line (extending Ref.~\cite{Gaiotto2013}) could settle this discrepancy.

For $\psi_{3/2}$ we find $1.932(11)$, $1.1\sigma$ below MC [$1.99(5)$] with a $4.5$ times smaller uncertainty. The spin-$2$ primary gives $\Delta_{t_+}=3.08(2)$, compared with $3.1(5)$ from MC~\cite{Billo2013} and $3.111$ at one loop~\cite{Gaiotto2013}. For the S-odd vector we find $\Delta_{D^o}=3.185(14)$, within $0.3\sigma$ of MC [$3.3(2)$ with a systematic error $0.3$]~\cite{Billo2013}. Since S-parity is emergent, $D^o$ is identified not by a quantum number but by its small $\Lambda_z$ matrix elements ($r\le0.09$) with all lower levels of its sector, an order of magnitude below those of descendant pairs (Sec.~\ref{sec:lambda}).

In the $j=0$ sector the DMRG levels at $2s+1=17$--$25$ continue the ED levels smoothly [Fig.~\ref{fig:extrap}(g,h)]. The even scalar rises slowly, from $2.262$ at $2s+1=13$ to $2.276$ at $2s+1=25$; the three fits give $2.290$, $2.291$, and $2.300$, so $\Delta_{\mathcal O_s}=2.295(11)$, $1.7\sigma$ above MC [$2.27(1)$]~\cite{Billo2013} . The pseudoscalar $p^o$, with $R=+1$ and no $\Lambda_z$ coupling to lower states, drifts downward, from $2.952$ to $2.886$ over the same sizes, and extrapolates to $\Delta_{p^o}=2.79(3)$, compared with $2.9(2)$ (MC~\cite{Billo2013}) and $2.92$ (bootstrap~\cite{Gaiotto2013}). The run at $2s+1=25$ is not yet converged: its levels moved by about $2\times10^{-3}$ in the last sweep, and a further change of that size would shift the three-size fits by up to $0.02$. The estimates do not, however, depend on this point: fits that omit $2s+1=25$, using $2s+1=17$ and $21$ alone or together with ED at $2s+1=13$, give $2.2956$--$2.2958$ for $\mathcal O_s$ and $2.792$--$2.794$ for $p^o$, and adding ED at $2s+1=11$ changes neither by more than $0.005$. The descendant $\partial_\parallel\mathcal O_s$ extrapolates to $3.31(2)$, one unit above $\Delta_{\mathcal O_s}$ within its error. Finally, the four higher-lying primaries resolved by DMRG are $\chi_{1/2}$ [$3.705(18)$], $\chi_{3/2}$ [$3.296(14)$], $\chi_2$ [$3.838(17)$], and $\chi_2'$ [$4.17(3)$], and the second vector is $\chi_1$ [$3.80(2)$], $4.5\%$ below the bootstrap estimate $3.98$~\cite{Gaiotto2013}. The level $\chi_2'$ may correspond to the S-odd spin-$2$ operators $t^o_\pm$, for which MC gives only the bound $\Delta\ge4.2(1)$~\cite{Billo2013}.

\section{Summary and outlook}
\label{sec:summary}
Multi-target DMRG extends fuzzy-sphere spectroscopy of the Ising twist line to $2s+1=25$ ($49$ orbitals), deep enough to reach the asymptotic scaling regime. With the displacement operator as the ruler and $\partial_\parallel D$ as a test, the finite-size corrections are dominated by the leading defect scalar $\mathcal O_s$, and the calibrated spectrum reproduces $\Delta_{\partial_\parallel D}=3$ to $0.4\%$. We obtain $\Delta_{\psi_{3/2}}$, $\Delta_{t_+}$, $\Delta_{D^o}$, and $\Delta_{p^o}$ with $4$--$23$ times smaller uncertainties than previous MC estimates, find $\Delta_\psi=0.912(4)$ and $\Delta_{\mathcal O_s}=2.295(11)$, $1.6\sigma$ below and $1.7\sigma$ above MC, respectively, and resolve four higher-lying primaries. An independent calibration with the bulk stress tensor on the untwisted sphere agrees with the displacement ruler in the continuum limit to within about $2\%$.

Several extensions are direct. Converging the $j=0$ sector at $2s+1=25$ and extending all sectors to $2s+1=29$ and beyond will firm up the extrapolations, in particular of $\mathcal O_s$ and $p^o$; resolving $R$ in this sector will give the $R=-1$ pseudoscalar $\tilde p$ [MC: $3.7(2)$~\cite{Billo2013}]; and scans of the bond dimension and of $h$ around $h_c$ will bound the remaining systematic errors. The converged MPS give access to defect OPE coefficients and the defect $g$-function~\cite{Zhou2024gfunction}. The same construction applies to other monodromy defects, and networks of twist lines joined at junctions are a natural next target for probing the categorical data of the $\Ztwo$ symmetry.

On the computational side, the cost grows mildly with system size: from $2s+1=17$ to $25$ the bond dimension grows by a factor $2$--$3$ while the sector dimension grows by a factor of $3\times10^4$, as expected from the area law on the sphere. For the bulk, exploiting the full non-Abelian $\mathrm{SO}(3)$ symmetry is a natural route to larger sizes~\cite{Hao2026}. A line defect leaves only rotations about the line and $R$, so for defects further gains must come from other directions, such as extrapolation in the bond dimension.

More broadly, twisted boundary conditions realize monodromy defects of any internal symmetry that acts on the fermion flavors, and multi-target DMRG removes the system-size bottleneck of ED. The combination therefore extends to defects in multi-flavor fuzzy-sphere models, such as the $\mathrm{SO}(5)$ and $\mathrm{Sp}(N)$ theories~\cite{Zhou2024SO5,ZhouHe2025}, whose Hilbert spaces put them beyond the reach of ED even without a defect.

\emph{Note added.}---While this work was being finalized, Refs.~\cite{TangZhu2026,Qian2026,TangZhu2026U1,LeeNaskar2026} appeared. Refs.~\cite{TangZhu2026,Qian2026} independently realize the $\Ztwo$ twist of the 3D Ising CFT on the fuzzy sphere by twisted boundary conditions, and the approach extends to other monodromies: the $\mathrm O(2)$ $\pi$-twist and three $\mathrm O(3)$ ones~\cite{TangZhu2026}, continuous families of $\mathrm O(2)$ and $U(1)$ twists~\cite{Qian2026,TangZhu2026U1}. Ref.~\cite{LeeNaskar2026} studies $\mathrm O(2)$ and $\mathrm O(3)$ monodromy defects and their bulk renormalization group flows. Where the results overlap they are consistent with ours, and the DMRG spectrum reported here reaches $2s+1=25$.

\begin{acknowledgments}
  The authors thank Zheng Zhou and Wei Zhu for invaluable discussions.
  This work was supported by National Key R\&D Program of China (Grant No. 2022YFA1403402), by National Natural Science Foundation of China (Grant No. 12374144), and by the Science and Technology Commission of Shanghai Municipality (Grant Nos. 24LZ1400100 and 23JC1400600).
  Y.Q. acknowledges the support of Shuguang Program of Shanghai Education Development Foundation and Shanghai Municipal Education Commission.
\end{acknowledgments}

\appendix

\section{Interaction matrix elements on the twisted grid}
\label{app:matrix}
The pseudopotential decomposition of the bulk interaction,
\begin{equation}
V_{m_1m_2m_3m_4}=\sum_{\ell}V_\ell\,
C^{2s-\ell,M}_{s,m_1;s,m_2}\,C^{2s-\ell,M}_{s,m_4;s,m_3},
\end{equation}
with $M=m_1+m_2$ and pair angular momentum $J=2s-\ell$, is continued to half-integer $m_i$ by writing the Clebsch--Gordan
coefficients in terms of $\Gamma$ functions. For the leading channel
$J=2s$,
\begin{multline}
C^{2s}_{m_1m_2}=\bigg[\frac{\Gamma(2s+1)^2}{\Gamma(4s+1)}\,\Gamma(2s-M+1)\,\Gamma(2s+M+1)\\
\times\prod_{i=1,2}\frac{1}{\Gamma(s-m_i+1)\,\Gamma(s+m_i+1)}\bigg]^{1/2}.
\label{eq:CG}
\end{multline}
For the contact interaction this continuation is exact. With the
orbitals~\eqref{eq:LLL}, the substitution $u=\sin^2(\theta/2)$ turns the overlap
integral into a Beta function,
\begin{multline}
\int d\Omega\,\Phi^*_{s,m_1}\Phi^*_{s,m_2}\Phi_{s,m_3}\Phi_{s,m_4}\\
=4\pi\,\delta_{M,m_3+m_4}\,
\frac{\Gamma(2s-M+1)\Gamma(2s+M+1)}{\Gamma(4s+2)}\prod_{i=1}^4\mathcal N_{s,m_i}\\
=\frac{(2s+1)^2}{4\pi(4s+1)}\,C^{2s}_{m_1m_2}C^{2s}_{m_4m_3},
\end{multline}
and the derivation uses only $s\pm m_i>-1$. It therefore holds unchanged on
the integer grid, on the half-integer grid and for mixed flavors, where $M$
is half-integer. We have also checked it numerically at $2s+1=11$ for integer,
half-integer and mixed index sets, with agreement to $10^{-15}$. On the
integer grid the continued coefficients reproduce
FuzzifiED~\cite{Zhou2025FuzzifiED} to machine precision.
Similarly, the subleading pseudopotential channel $J=2s-1$ ($\ell=1$) is
continued analytically in an identical manner via $\Gamma$ functions,
matching the exact overlap integrals and reproducing the FuzzifiED implementation
to $10^{-15}$. On the half-integer grid the orbitals do not form $\mathrm{SO}(3)$
multiplets, so for $\ell\ge1$ this continuation defines the interaction on the
twisted grid rather than following from rotation symmetry; it is fixed by
requiring that the matrix elements equal the real-space overlap integrals of
the same kernel.

\section{Multi-target DMRG: algorithm and cost}
\label{app:algo}
\textit{Block Lanczos update.} At each two-site step the bundled MPS supplies
$K$ orthonormal centers, collected as the columns of a matrix $Q_0$ of size
$d_{\rm eff}\times K$, where $d_{\rm eff}$ is the dimension of the two-site
effective space. The effective Hamiltonian $H_{\rm eff}$ is then reduced by a
block Lanczos recursion~\cite{Baker2024,Hao2026}. With $Q_{-1}=0$, one computes
for $k=0,1,\dots,p-1$
\begin{align}
A_k&=Q_k^\dagger H_{\rm eff}Q_k,\\
\tilde Q_{k+1}&=H_{\rm eff}Q_k-Q_kA_k-Q_{k-1}B_k^\dagger,\\
\tilde Q_{k+1}&=Q_{k+1}B_{k+1},
\end{align}
where the last step is a QR (or SVD polar~\cite{Hao2026}) reorthonormalization, so that
$Q_{k+1}^\dagger Q_{k+1}=\mathbb 1_K$ and
$B_{k+1}=Q_{k+1}^\dagger H_{\rm eff}Q_k$. In the basis
$P=[Q_0,\dots,Q_p]$, $H_{\rm eff}$ becomes block tridiagonal, with $A_k$ on
the diagonal and $B_k$ next to it. This $K(p+1)$-dimensional matrix is
diagonalized exactly. Its $K$ lowest eigenvectors $c_k$ give the new centers
$\Theta_k=Pc_k$, which are orthonormal before bond truncation. For $K=1$ the procedure reduces to
the ordinary Lanczos step of single-state DMRG. The new bond basis consists
of the $\chi$ leading eigenvectors of
$\rho=K^{-1}\sum_k\mathrm{Tr}_{\rm env}\,\Theta_k\Theta_k^\dagger$, onto which all $K$
centers are projected; at the end of each full sweep, the $K\times K$ Gram matrix $G_{ij}=\langle\psi_i|\psi_j\rangle$ is computed to whiten the bundle via $G^{-1/2}$, followed by a $K\times K$ Ritz rotation of $H$ that restores exact global orthonormality.

\textit{Why it scales.} Every operation acts on the $K$ centers at once, so
the cost of an update is that of $K$ applications of $H_{\rm eff}$, and the
cost of a sweep scales as $N_{\rm orb}K\chi^3D_W$, where $D_W$ is the bond
dimension of the matrix-product operator. Two features make the method
suited to large spheres. First, the bond dimension is set by entanglement.
The LLL orbitals ordered by $m$ are localized on circles of latitude, a cut
of the chain is a cut of the sphere along such a circle, and the
entanglement follows an area law with boundary length $\propto
R\propto\sqrt{N_{\rm orb}}$~\cite{Hao2026}. The required $\chi$ grows as
$e^{a\sqrt{N_{\rm orb}}}$, compared with the $e^{bN_{\rm orb}}$ growth of the
Hilbert space. Second, the $K$ targets are found in one calculation. In the
projector method, by contrast, $K$ separate runs are needed, each run
inherits the errors of the earlier ones, and the penalty weight must be
tuned. Here the $K$ states are exact Ritz vectors of the same projected
Hamiltonian and are orthonormal by construction. The cost of this is a larger
$\chi$. In our runs the $K=4$ bundles reach about five times the bond
dimension of the $K=1$ ground-state run (a different sector; Table~\ref{tab:conv}).

\begin{figure*}[tp]
\centering
\includegraphics[width=0.82\textwidth]{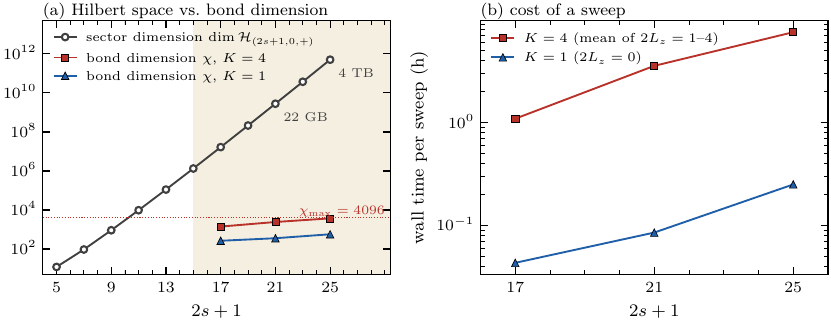}
\caption{Cost of the calculation, showing why the multi-target DMRG reaches
sizes that ED cannot. (a)~Exact dimension of the ground-state sector
$(2s+1,0,+)$ (open circles), with the memory of one double-precision vector
indicated at $2s+1=21,25$. For comparison, the largest bond dimension reached
in the $K=4$ runs (squares, maximum over $2L_z=1$--$4$) and in the $K=1$
ground-state run (triangles). The Hilbert space grows exponentially in the
number of orbitals; the bond dimension grows slowly, as expected from the
area law on the sphere. The dotted line is the cap $\chi_{\max}=4096$, which
the main runs do not reach. (b)~Wall time of the last sweep, averaged over the
four $K=4$ sectors (squares), and for the $K=1$ sector (triangles). Shaded:
DMRG range.}
\label{fig:cost}
\end{figure*}

Figure~\ref{fig:cost} quantifies this. From $2s+1=17$ to $2s+1=25$ the dimension
of the ground-state sector, Table~\ref{tab:dims}, grows by a factor of $3\times10^4$. The
memory needed to store a single vector in double precision rises from
$22$~GB at $2s+1=21$ to $4$~TB at $2s+1=25$, which puts $2s+1=25$ out of reach of ED.
Over the same range the bond dimension of the $K=4$ runs grows by a factor
of $2.5$--$2.7$ and stays below the cap $4096$; that of the $K=1$ run grows by
$2.2$.
The wall time per sweep grows by a factor of $7$ for $K=4$ and $6$ for
$K=1$.

\begin{table*}[tp]
\caption{Dimension of the sector $(\Ne,2L_z,(-1)^{\Nm})=(2s+1,0,+)$ that
contains the defect ground state, counted exactly on the
orbitals~\eqref{eq:orbitals}.}
\label{tab:dims}
\begin{ruledtabular}
\begin{tabular}{lccccccc}
$2s+1$ & 7 & 9 & 11 & 13 & 17 & 21 & 25\\
total orbitals $4s+1$ & 13 & 17 & 21 & 25 & 33 & 41 & 49\\
dimension & $94$ & $910$ & $9686$ & $1.1\times10^5$ & $1.7\times10^7$ & $2.8\times10^{9}$ & $5.0\times10^{11}$\\
\end{tabular}
\end{ruledtabular}
\end{table*}

\section{DMRG convergence}
\label{app:conv}
\begin{figure*}[tp]
\centering
\includegraphics[width=\textwidth]{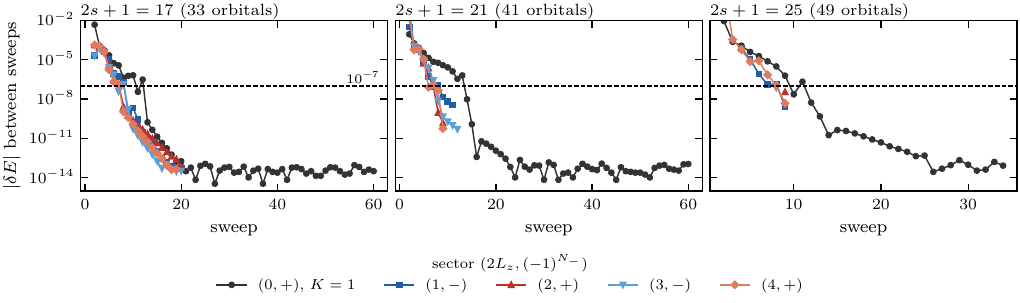}
\caption{Convergence of the main multi-target DMRG runs (the $j=0$ runs with
$K=4$ and the untwisted runs are summarized in Table~\ref{tab:conv}). Each panel is one system size;
each curve is one symmetry sector $(2L_z,(-1)^{\Nm})$ and shows the largest
change $|\delta E|$ among the $K$ target energies between consecutive
sweeps. The vacuum sector $(0,+)$ targets only $\one$ ($K=1$); the others
target $K=4$ states. The density-matrix perturbation is switched off from
sweep $7$ on. The dashed line is the stopping criterion $10^{-7}$. All runs
reach it, including those at $2s+1=25$, where the twist sectors use bond
dimensions up to $3703$. The plateau at $10^{-14}$ for $2s+1=17$ is the
floating-point limit. The sweep-to-sweep changes are far below the
finite-size drift of the levels.}
\label{fig:conv}
\end{figure*}

Figure~\ref{fig:conv} shows the energy change between consecutive sweeps for
all runs, and Table~\ref{tab:conv} lists the largest bond dimension $\chi$
and the energy change $\delta E$ between the last two sweeps. The sectors are
labeled by $(2L_z,(-1)^{\Nm})$. In the main runs the $2L_z=0$ sector contains only the defect
ground state $\one$ ($K=1$); the others contain $K=4$ targets. At $2s+1=25$ the
largest bond dimension is $3703$ and the discarded weights are about
$10^{-7}$. The energy change per
sweep is below $10^{-7}$, four orders of magnitude below the finite-size
drift of the levels between $2s+1=21$ and $25$. The sweep-to-sweep changes are therefore
well below the extrapolation and calibration uncertainties; a bond-dimension
extrapolation at the largest sizes is left for future work.

\textit{The $j=0$ sector and the truncation error.} The excited states of the
$2L_z=0$ sector are computed in separate runs with $K=4$ targets, $\one$,
$\mathcal O_s$, $p^o$, and $\partial_\parallel\mathcal O_s$ (Table~\ref{tab:conv}).
At $2s+1=17$ and $21$ these runs converge with $\chi=936$ and $1575$. At $2s+1=25$
the run has completed two sweeps at the cap $\chi=4096$; the second sweep
lowered the target energies by $(2$--$10)\times10^{-3}$ and the calibrated $j=0$
levels by $(1.9$--$3.7)\times10^{-3}$. Its ground-state energy, however, already
lies $4.6\times10^{-4}$ below that of the $K=1$ vacuum run, continuing the
differences $1.7\times10^{-4}$ and $3.0\times10^{-4}$ found at $2s+1=17$ and $21$.
These differences arise because the $K=4$ bundle needs a larger bond
dimension than a single state, so its lowest state is represented more
accurately. They measure the truncation error of the $K=1$ vacuum energy, which
is much larger than its sweep-to-sweep change. An error of this size in one
energy shifts the calibrated levels by about $10^{-4}$. To compute both energies
at the same truncation, we measure the $j=0$ levels from the ground state of
their own bundle.

\textit{Untwisted sphere.} For the bulk ruler of Sec.~\ref{sec:calibration} we
solve the untwisted model at the same couplings for the vacuum and for the
stress tensor, the lowest $\Ztwo$-even state at $L_z=2$, each with $K=1$
(Table~\ref{tab:conv}, lower block). The gap $\Delta E_T$ falls from $10.444$ at
$2s+1=17$ to $8.579$ at $2s+1=25$.

\textit{Benchmark against ED.} The DMRG code was tested against ED of the
same twisted Hamiltonian, with the parameters of the production runs
($V_0=4.75$, $V_1=1$, $h=3.16$, $\Ne=2s+1$), four targets per sector and
random initial MPS that contain no information from ED. At $2s+1=5$ all nine
sectors $2L_z=-4,\dots,4$ were solved ($\chi\le32$); at $2s+1=7$ the same nine
sectors and, with a different seed, the five sectors $2L_z=0,\dots,4$
($\chi\le128$). In these $23$ sector runs ($92$ states) the largest energy
deviation from ED is $1.1\times10^{-13}$, the largest normalized residual
$\|(H-E)\Psi\|/|E|$ is $1.0\times10^{-12}$, and the calibrated ratios
$\hD$ agree with ED to $2\times10^{-14}$. At $2s+1=9$ the sectors $(0,+)$
with $K=2$ and $(2,+)$, $(1,-)$ with $K=1$ converge to ED within
$5\times10^{-14}$ once $\chi$ is two to five times the bundle capacity $C$, the
largest bond dimension needed to represent the $K$ targets exactly, counted
sector by sector at each bond; the extra room is used during the perturbed
sweeps. ED and
DMRG are built from the same interaction terms, so this test checks the
solver, the conversion to an MPO and the fermion signs, not the derivation of
the interaction itself, which is checked separately in Appendix~\ref{app:matrix}.

\begin{table*}[tp]
\caption{Convergence of the DMRG runs: largest bond dimension $\chi$ and
energy change $\delta E$ between the last two sweeps. Upper block: sectors
$(2L_z,(-1)^{\Nm})$ of the twisted sphere. The $j=0$ run with $K=4$ has
completed two sweeps at $2s+1=25$ (preliminary).
Lower block: the two states of the untwisted sphere that set the bulk ruler,
the vacuum and the stress tensor $T$ (lowest $\Ztwo$-even state at $L_z=2$),
each with $K=1$.}
\label{tab:conv}
\begin{ruledtabular}
\begin{tabular}{lcccccc}
 & \multicolumn{2}{c}{$2s+1=17$} & \multicolumn{2}{c}{$2s+1=21$} & \multicolumn{2}{c}{$2s+1=25$}\\
sector & $\chi$ & $\delta E$ & $\chi$ & $\delta E$ & $\chi$ & $\delta E$\\
\hline
$(0,+)$, $K=1$ & 263 & $3\times10^{-14}$ & 355 & $1\times10^{-13}$ & 567 & $9\times10^{-14}$\\
$(0,+)$, $K=4$ & 936 & $5\times10^{-14}$ & 1575 & $1\times10^{-8}$ & 4096 & $1\times10^{-2}$\\
$(1,-)$, $K=4$ & 960 & $6\times10^{-14}$ & 1594 & $4\times10^{-9}$ & 2381 & $3\times10^{-9}$\\
$(2,+)$, $K=4$ & 1162 & $6\times10^{-14}$ & 2012 & $2\times10^{-10}$ & 3086 & $4\times10^{-8}$\\
$(3,-)$, $K=4$ & 1012 & $3\times10^{-14}$ & 1671 & $5\times10^{-11}$ & 2523 & $7\times10^{-8}$\\
$(4,+)$, $K=4$ & 1405 & $4\times10^{-14}$ & 2409 & $6\times10^{-11}$ & 3703 & $5\times10^{-9}$\\
\hline
bulk $\one$, $K=1$ & 253 & $2\times10^{-14}$ & 399 & $1\times10^{-12}$ & 597 & $8\times10^{-11}$\\
bulk $T$, $K=1$ & 526 & $2\times10^{-9}$ & 830 & $5\times10^{-10}$ & 1218 & $3\times10^{-8}$\\
\end{tabular}
\end{ruledtabular}
\end{table*}

\section{Size dependence of the levels}
\label{app:size}
\begin{figure*}[tp]
\centering
\includegraphics[width=\textwidth]{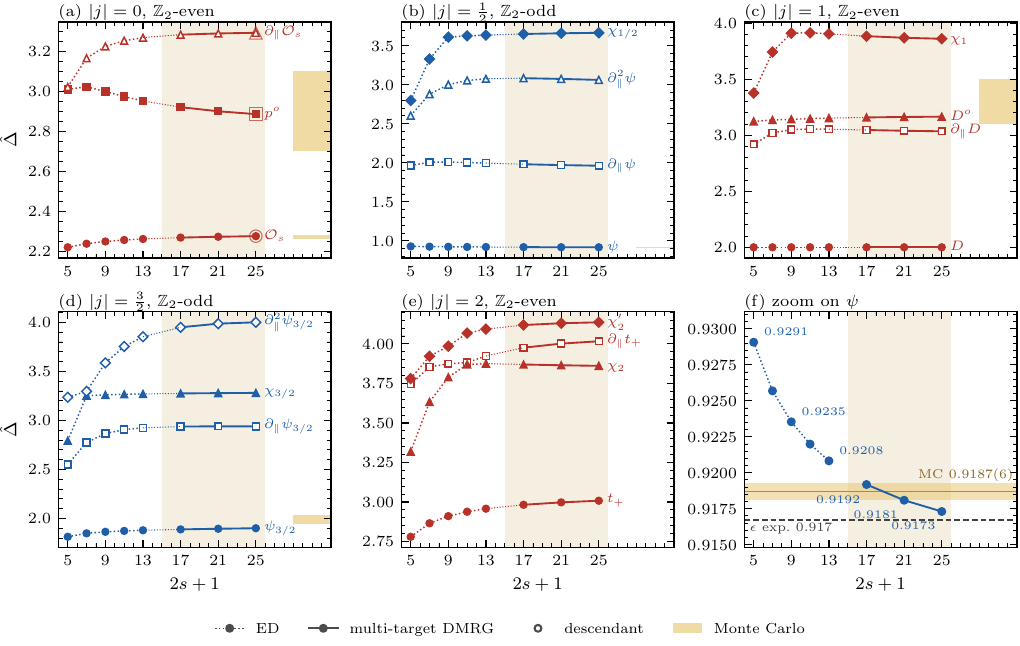}
\caption{Calibrated dimensions versus $2s+1$, one panel per symmetry sector
$(|j|,\Ztwo)$, all computed in the twisted system. Dotted lines with symbols:
ED ($2s+1=5$--$13$). Solid lines: multi-target DMRG ($2s+1=17$--$25$, shaded). Filled
symbols are primaries and open symbols line descendants; each curve is
labeled at its largest size. Shaded bars at the right edge are MC values with
their errors. (a)~$|j|=0$, $\Ztwo$-even: $\mathcal O_s$, $p^o$ and
$\partial_\parallel\mathcal O_s$, from separate $K=4$ runs of the $2L_z=0$
sector; the large ring marks the preliminary point
(Appendix~\ref{app:conv}).
(b)--(e)~The sectors $|j|=\tfrac12,1,\tfrac32,2$ computed with DMRG. In (c)
$D$ is fixed to $2$ by the calibration. (f)~Enlarged view of $\psi$, with
values at selected sizes, the MC band $0.9187(6)$ and the one-loop
$\epsilon$-expansion value $0.917$ (dashed). The levels change smoothly and
do not cross in the DMRG range, which is what makes the identifications of the
main text unambiguous.}
\label{fig:Ds}
\end{figure*}

Figure~\ref{fig:Ds} shows every calibrated level as a function of $2s+1$, with ED
for $2s+1=5$--$13$ and DMRG for $2s+1=17,21,25$. It serves two purposes. It shows
that the ED and DMRG data join smoothly, so the two methods describe the same
system. It also shows that the level assignments are stable: between $2s+1=17$
and $2s+1=25$ no two levels of the same sector cross, and every level moves by
less than $0.06$. Below $2s+1=17$ the picture is less clean. In the
$|j|=\tfrac32$ sector, for example, $\partial_\parallel^2\psi_{3/2}$ and
$\chi_{3/2}$ cross in the ED range. Panel (f) enlarges $\psi$ and shows its
slow downward drift through the MC band, discussed in the main text.

\section{Convergence of descendant spacings}
\label{app:spacings}
\begin{figure*}[tp]
\centering
\includegraphics[width=\textwidth]{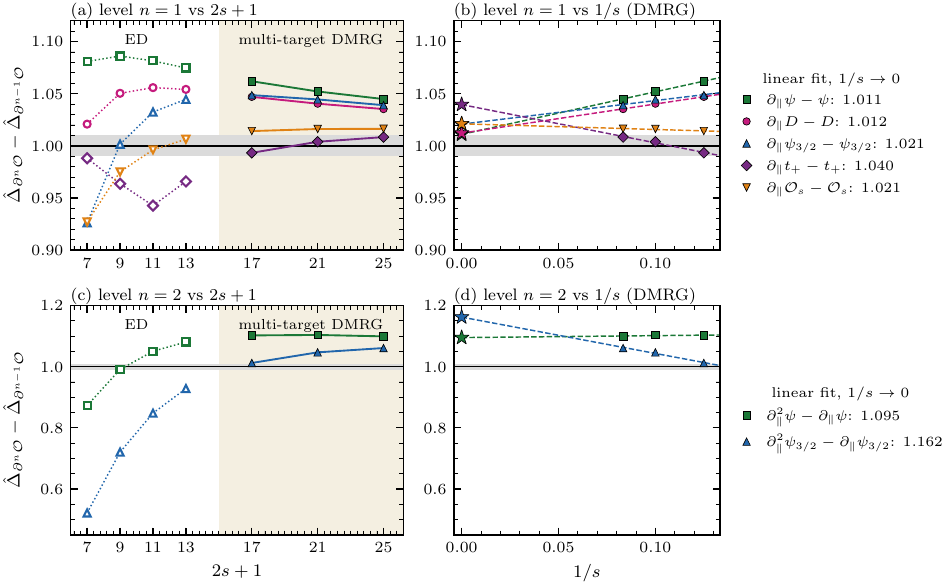}
\caption{Consecutive descendant steps
$\hD_{\partial^n\calO}-\hD_{\partial^{n-1}\calO}$, which equal $1$ in the
continuum (black line; the gray band marks $\pm1\%$). (a,b)~Level $n=1$ for
the towers of $\psi$ (squares), $D$ (circles), $\psi_{3/2}$ (triangles),
$t_+$ (diamonds) and $\mathcal O_s$ (down triangles); (c,d)~level $n=2$ for $\psi$ and $\psi_{3/2}$.
(a,c)~Versus $2s+1$: ED ($2s+1=7$--$13$, open symbols, dotted lines) and
multi-target DMRG ($2s+1=17$--$25$; filled symbols, shaded). (b,d)~DMRG data
versus $1/s$ with linear fits (dashed); the stars at $1/s=0$ are the
extrapolated steps, whose values are listed in the legends. The level-one
steps of the three cleanest towers extrapolate to $1$ within $1$--$2\%$, a
direct check of conformal symmetry along the line. The level-two steps and
the $t_+$ tower converge more slowly, for the reasons given in the text.}
\label{fig:spacings}
\end{figure*}

A line primary $\calO$ and its descendants $\partial_\parallel^n\calO$ form
an $\mathrm{SL}(2,\mathbb R)$ multiplet, so in the continuum the consecutive
steps $\hD_{\partial^n\calO}-\hD_{\partial^{n-1}\calO}$ equal $1$ exactly.
These steps are fixed by symmetry and need no external input. How fast they
approach $1$ as $2s+1$ grows is therefore a test of the regularization, the
calibration and the state assignments that is independent of any comparison
with MC. Figure~\ref{fig:spacings} shows the steps for the five towers with a
resolved level-one descendant ($\psi$, $D$, $\psi_{3/2}$, $t_+$, $\mathcal O_s$) and the two
towers with a resolved level-two descendant ($\psi$, $\psi_{3/2}$).

At level one the ED data ($2s+1\le13$) scatter by up to $8\%$ and are not
monotonic, whereas the DMRG data ($2s+1=17$--$25$) change smoothly and
monotonically. For $\psi$, $D$ and $\psi_{3/2}$ the step decreases from
$1.047$--$1.062$ at $2s+1=17$ to $1.035$--$1.045$ at $2s+1=25$, and linear fits in
$1/s$ give $1.011$, $1.012$ and $1.021$. The step of $\partial_\parallel D$
is the protected combination $\Delta_{\partial D}-\Delta_D=1$ and is the most
direct check of the displacement ruler. The $t_+$ step lies within about $1\%$ of
unity at all DMRG sizes but drifts slowly upward, and its linear
extrapolation, $1.040$, should not be overinterpreted, since the level-one
state of the $t_+$ tower is nearly degenerate with the primary $\chi_2$ and
the two mix at finite size. The $j=0$ step of $\mathcal O_s$ is nearly flat ($1.014$--$1.016$) and extrapolates linearly to
$1.021$; the other two forms of Appendix~\ref{app:budget} give $1.004$ and $1.024$.

At level two the steps remain $6$--$10\%$ above unity at $2s+1=25$. For $\psi$
the step is nearly independent of size ($1.102$ to $1.099$), and for
$\psi_{3/2}$ it increases from $1.012$ at $2s+1=17$ to $1.061$ at $2s+1=25$, after
the level crossing with $\chi_{3/2}$ visible in the ED range. The
level-two descendants are the highest states of their bundles, and their
finite-size corrections are larger and not linear in $1/s$; we therefore do
not use them in the extrapolations of the main text.

\section{Error budget and extrapolation form}
\label{app:budget}
\begin{table*}[t]
\caption{Infinite-size values of selected levels from different
extrapolation forms, all fitted to the three DMRG points $2s+1=17$--$25$; the
quadratic and free-$\omega$ entries are exact three-point solutions. The first three forms enter the error budget. The last column is the
final estimate.}
\label{tab:forms}
\begin{ruledtabular}
\begin{tabular}{lccccccc}
level & $1/s$ & $1/s$, quadr. & $s^{-0.635}$ & $s^{-1/2}$ & $s^{-0.415}$ & free $\omega$ & final\\
\hline
$\psi$ & $0.9136$ & $0.9127$ & $0.9109$ & $0.9089$ & $0.9070$ & $0.9111$ & $0.912(4)$\\
$\psi_{3/2}$ & $1.9236$ & $1.9367$ & $1.9403$ & $1.9526$ & $1.9646$ & unstable & $1.932(11)$\\
$D^o$ & $3.1797$ & $3.1842$ & $3.1894$ & $3.1966$ & $3.2035$ & $3.1972$ & $3.185(14)$\\
$t_+$ & $3.0594$ & $3.0643$ & $3.0958$ & $3.1228$ & $3.1488$ & $3.0707$ & $3.08(2)$\\
$\mathcal O_s$ & $2.2902$ & $2.2907$ & $2.3003$ & $2.3078$ & $2.3150$ & $2.2913$ & $2.295(11)$\\
$p^o$ & $2.8165$ & $2.8080$ & $2.7672$ & $2.7307$ & $2.6955$ & $2.7957$ & $2.79(3)$\\
$\partial_\parallel D$ & $3.0121$ & $3.0001$ & $2.9955$ & $2.9833$ & $2.9714$ & $2.8896$ & $3.004(15)$\\
\end{tabular}
\end{ruledtabular}
\end{table*}

\textit{Origin of the exponent.} At finite size the fuzzy-sphere Hamiltonian
flows to the fixed point plus every irrelevant operator allowed by symmetry,
$H=H_*+\sum_i g_i\int\mathcal O_i$. On a sphere of radius $R$ an operator of
dimension $\Delta_i$ integrated over a region of dimension $d_i$ ($d_i=3$ in the
bulk, $d_i=1$ on the line) shifts the levels by $g_iR^{-\Delta_i+d_i-1}$ in
first-order perturbation theory, while the level spacing is $1/R$. The
calibrated dimensions therefore receive corrections $\propto
R^{-(\Delta_i-d_i)}$. The $N_{\rm orb}\propto s$ orbitals cover the sphere at
fixed density, so $R\propto\sqrt s$ in units of the magnetic length, which
gives Eq.~\eqref{eq:omega} with exponents $\omega_i=(\Delta_i-d_i)/2$. The
$\Ztwo$-even, rotation-invariant bulk operators are $\epsilon$, which is tuned
away by $h$, then $T_{\mu\nu}$, whose integral is a conserved charge, and
$\epsilon'$ with $\omega=0.415$. On the line the lowest $\Ztwo$-even scalar
other than the identity is $\mathcal O_s$ with $\Delta_{\mathcal O_s}=2.27(1)$, so
$\omega_{\mathcal O_s}=0.635$. Higher operators and second-order terms give
$s^{-1}$, $s^{-0.83}$ (second order in $\epsilon'$), $s^{-1.27}$ (second order
in $\mathcal O_s$) and beyond. With our own value $\Delta_{\mathcal O_s}=2.295$
the defect exponent is $0.648$ instead of $0.635$; this changes every
extrapolated value in Table~\ref{tab:results} by at most $0.0016$, less than
$5\%$ of its error.

\textit{Test on $\partial_\parallel D$.} Since $\Delta_{\partial D}=3$ is exact,
$\hD_{\partial D}(s)-3$ isolates the corrections. Fitting the three DMRG points
with $\Delta=3$ fixed gives an rms residual of $1.3\times10^{-4}$ for
$b\,s^{-0.415}+c\,s^{-0.635}$ ($b=-0.033$, $c=0.228$), $4.8\times10^{-4}$ for
$c\,s^{-0.635}$ alone, $2.0\times10^{-3}$ for $c/s$ and $2.0\times10^{-3}$ for
$b\,s^{-0.415}$ alone. The defect term thus dominates and the bulk term is a
small correction, as expected for the spectrum of a line. A fit
$3+c\,s^{-\omega}$ with $\Delta=3$ fixed and free $\omega$ gives
$\omega=0.70\pm0.02$. The effective exponent lies between
$\omega_{\mathcal O_s}=0.635$ and $1$, which indicates that the subleading
terms listed above contribute at these sizes. Three points do not determine
them separately: with $\Delta$ also free the fit drifts to $\omega\approx0.2$ and misses $\Delta_{\partial D}=3$ by $3.7\%$, and it is unstable for operators with
smaller drift. We therefore use three two- or three-parameter forms that
bracket the effective exponent, linear and quadratic in $1/s$ and linear in
$s^{-0.635}$, and take their spread as the extrapolation error.
Table~\ref{tab:forms} shows that forms outside this bracket ($s^{-1/2}$,
$s^{-0.415}$) miss $\Delta_{\partial D}=3$ by $0.6\%$ and $1.0\%$.

\textit{Error budget.} For each operator the central value is the midpoint of
the three fits, and the extrapolation error $\sigma_{\rm fit}$ is half their spread. The
calibration error is the largest deviation of the three $\partial_\parallel D$
fits from $3$, $0.40\%$. A residual error of the ruler rescales all
levels by a common factor, so it is applied as $\sigma_{\rm cal}=0.0040\,\Delta$. The two are added in
quadrature, $\sigma=(\sigma_{\rm fit}^2+\sigma_{\rm cal}^2)^{1/2}$. DMRG truncation errors are not included. The comparison of the $K=1$ and $K=4$ vacuum energies (Appendix~\ref{app:conv}) puts them at a few $10^{-4}$ per energy, which shifts the calibrated levels by about $10^{-4}$, more than an order of magnitude below $\sigma$. This uncertainty budget quantifies the functional-form and ruler uncertainties at fixed $h=3.16$; potential systematic shifts from the small finite-size drift of the bulk critical point ($h_c\simeq3.153$--$3.16$~\cite{Zhu2023,Hao2026}) are not included in $\sigma$. For $\psi$ the result $0.912(4)$ is dominated by the calibration
error; for the higher levels the two contributions are comparable. The
values in Table~\ref{tab:results} follow from this rule. With three DMRG sizes
the quadratic fit is exact. For the $j=0$ levels $\sigma$ does not include the
remaining convergence of the point at $2s+1=25$, which is discussed in Sec.~\ref{sec:primaries}.

\bibliography{refs_prl}

\end{document}